\documentclass[10pt,letterpaper,twocolumn]{article} 

\usepackage{ol2SmallCaptions}
\usepackage{comment}
\usepackage[draft]{hyperref}
\usepackage{amsmath}
\usepackage{subfigure}    
\usepackage{empheq}  
\usepackage{graphicx}	
\usepackage{epstopdf}

\begin{document}

\twocolumn[ 

\title{Tunable coupled-mode dispersion compensation and its application to on-chip resonant four-wave mixing} \vspace{-10pt}


\author{Cale M. Gentry$^{1*}$, Xiaoge Zeng$^2$, and Milo\v{s} A. Popovi\'{c}$^{1\dagger}$}
\address{
$^1$Department of Electrical, Computer, and Energy Engineering, University of Colorado Boulder, Colorado, 80309, USA\\
$^2$Department of Physics, University of Colorado Boulder, Colorado, 80309, USA\\
$^*$cale.gentry@colorado.edu, $^\dagger$milos.popovic@colorado.edu
}\vspace{-10pt}

\begin{abstract}
We propose and demonstrate localized mode coupling as a viable dispersion engineering technique for phase-matched resonant four-wave mixing (FWM).  We demonstrate a dual-cavity resonant structure that employs coupling-induced frequency splitting at one of three resonances to compensate for cavity dispersion, enabling phase-matching.  Coupling strength is controlled by thermal tuning of one cavity enabling active control of the resonant frequency-matching.  In a fabricated silicon microresonator, we show an 8\,dB enhancement of seeded FWM efficiency over the non-compensated state.  The measured four-wave mixing has a peak wavelength conversion efficiency of $-37.9$\,dB across a free spectral range (FSR) of 3.334\,THz ($\sim$27\,nm). Enabled by strong counteraction of dispersion, this FSR is, to our knowledge, the largest in silicon to demonstrate FWM to date.  This form of mode-coupling-based, active dispersion compensation can be beneficial for many FWM-based devices including wavelength converters, parametric amplifiers, and widely detuned correlated photon-pair sources. Apart from compensating intrinsic dispersion, the proposed mechanism can alternatively be utilized in an otherwise dispersionless resonator to counteract the detuning effect of self and cross phase modulation on the pump resonance during FWM, thereby addressing a fundamental issue in the performance of light sources such as broadband optical frequency combs.
\end{abstract}

\vspace{-15pt}
 ] 

\noindent On-chip four-wave mixing-based devices have been the subject of much research lately for applications such as wavelength converters \cite{Turner08}, octave spanning and phase-locked frequency combs \cite{Del'Haye11,Li12}, and quantum-entangled biphoton sources \cite{Clemmen09}.  Four-wave mixing (FWM) is a third-order nonlinear process arising from the $\chi^{(3)}$ susceptibility where two pump photons are parametrically converted to a signal-idler pair while conserving energy and momentum. Resonant enhancement of both the pump and signal/idler modes greatly improves FWM efficiency \cite{Absil00}.  Silicon has attracted interest as a nonlinear material due to a Kerr coefficient that is over 100 times that of silica \cite{Lin07} and the well developed fabrication of high quality factor (Q) optical resonators with small mode volumes.
\begin{figure}[b!]
    \centering
    \vspace{-25pt}
    \includegraphics[width= \columnwidth]{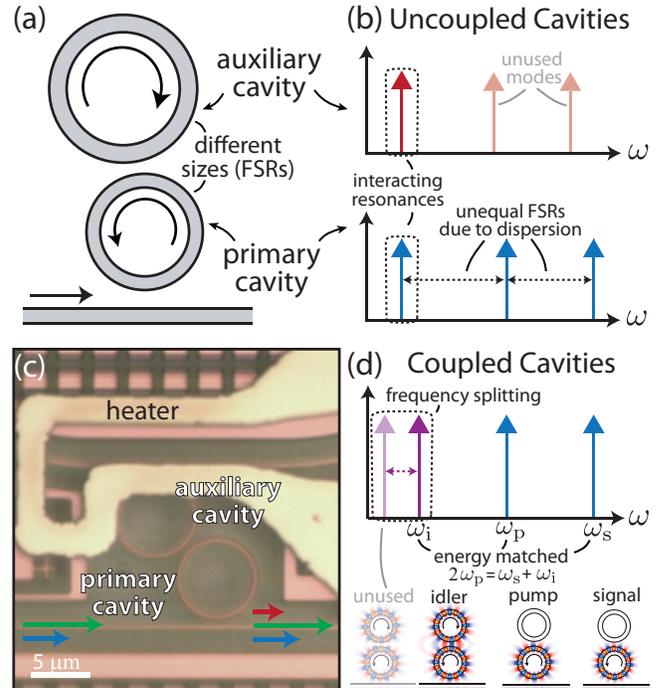}
    \vspace{-25pt}
  \caption{(a) Proposed resonator geometry of two differently sized coupled cavities with corresponding uncoupled resonance frequencies (b) that are not energy matched due to dispersion. (c) Optical micrograph of fabricated device. (d) Coupling results in frequency splitting at a single FSR allowing energy matching between three resonances \label{fig:fig1}}
\end{figure}

Aside from high Q and small modal volume, phase matching is critical to efficient resonant FWM.  In the case of a degenerate pump, three interacting wavelengths are present (pump, signal and idler).  Since cavity modes are intrinsically momentum matched \cite{Turner08}, efficient FWM is achieved when the signal, idler, and pump are on-resonance with the three modes.  The problem of efficient FWM is then reduced to designing for photon energy matching, i.e. for resonant modes equally spaced in frequency.  Adjacent longitudinal resonances of a single cavity, spaced by one free spectral range (FSR), are most commonly utilized for FWM due to device simplicity. Here, material and waveguide dispersion can lead to unequal FSRs and a serious reduction in FWM efficiency. Careful engineering of dispersion through resonator dimensions and pump wavelength is possible \cite{Turner08, Jiang12} but can impose severe restrictions on wavelength, mode volume, and FSR while conceding unnecessary fabrication complexity such as device layer thicknesses that require specialized fabrication techniques or the inclusion of additional materials \cite{Riemensberger12}. This method may also result in multiple pairs of FSRs being energy matched symmetrically around the pump, producing a comb of signal-idler wavelength pairs \cite{Jiang12} which can be disadvantageous for applications such as heralded single photon sources where additional signal-idler pairs result in inefficiencies.  Another promising solution is the use of the triply split resonance modes from a so-called `photonic molecule' consisting of three coupled optical cavities \cite{Azzini13, Zeng14}. The difficulty of obtaining resonant splitting greater than 10\,nm without sacrificing loss Q, due to coupler losses \cite{Zeng14}, suggests a need for other dispersion compensating methods.
 
Integrated optics has proven to be a platform where various cavity coupling strategies have provided new degrees of freedom in design, for example allowing the design of square yet dispersionless filters \cite{Popovic07}, dispersionless resonant switching \cite{Popovic07b}, linewidth-unlimited modulators \cite{Popovic10}, asymmetric-response-based high spectral efficiency multiplexers \cite{Wade13}, optomechanically coupled resonators for tailoring potentials through optical forces \cite{Rakich07}, and ultra-widely tunable single frequency lasers through imaginary, or radiative, coupling \cite{Gentry14}.

In this paper, we propose and demonstrate mode coupling as a strategy for resonance control and dispersion compensation to enable phase-matched FWM.  We demonstrate a dispersion compensating device that is one of the simplest embodiments of this idea, and compensates dispersive frequency offset of a single resonance.  The device consists of two coupled cavities with unequal FSRs referred to as the `primary' and `auxiliary' cavities as illustrated in Fig.~\ref{fig:fig1}.  The general principle is to use frequency splitting due to mode coupling at a single longitudinal resonance of the primary cavity to compensate for the inherent frequency mismatch caused by dispersion [Fig.~\ref{fig:fig1}(b)].  With strong enough coupling (frequency splitting), locally, a 4-way resonant system is created in which three resonances are energy matched and therefore suitable for FWM.  

We investigate the effect of unequally spaced resonances on FWM efficiency through a standard coupling of modes in time (CMT) model, which is consistent with the coupling of modes in space picture \cite{Absil00} commonly used for single rings. Here, the advantage of CMT is that it directly incorporates the interaction of resonant modes (rather than simply waveguide modes), and naturally supports multiple-cavity resonators with distributed super-modes in simple form \cite{ZengArxiv}. The dynamic equations for the pump, signal, and idler mode energy amplitudes ($a_{p}$, $a_{s}$, and $a_{i}$ respectively) in the resonator, in the undepleted pump and seed-signal approximation, are
\begin{align}
{d \over{dt}} a_p &= (j\omega_{p,o}-r_{p,o}-r_{p,\mathrm{ext}}) a_p -j \sqrt{2 r_{p,\mathrm{ext}}} s_{p,+} \nonumber \\
{d \over{dt}} a_s &= (j\omega_{s,o}-r_{s,o}-r_{s,\mathrm{ext}}) a_s -j \sqrt{2 r_{s,\mathrm{ext}}} s_{s,+} \label{eqn:cmt1} \\
{d \over{dt}} a_i &= (j\omega_{i,o}-r_{i,o}-r_{i,\mathrm{ext}}) a_i - j \omega_i \beta_{\mathrm{fwm},i} a_p^2 a_s^*  \nonumber 
\end{align}
\vspace{-15pt}
\begin{align}
s_{i,-} &= -j \sqrt{2 r_{i,\mathrm{ext}}} a_i   \label{eqn:cmt2} 
\end{align}
Here, $\omega_{p,o}$, $\omega_{s,o}$ and $\omega_{i,o}$ are the resonant frequencies, $r_{p,o}$, $r_{s,o}$ and $r_{i,o}$ are the decay rates due to linear intrinsic loss, and $r_{p,\mathrm{ext}}$, $r_{s,\mathrm{ext}}$ and $r_{i,\mathrm{ext}}$ are the decay rates due to external coupling to the waveguide bus for the pump, signal, and idler resonator modes respectively.  $s_{p,+}$ and $s_{s,+}$ are the power-normalized wave amplitude inputs in the waveguide for the pump and signal modes respectively and $s_{i,-}$ is the output wave amplitude for the idler mode.  The FWM coefficient $\beta_{\mathrm{fwm},i}$ contains the nonlinear properties of the material along with the modal field overlap as described in \cite{ZengArxiv}. As in \cite{Absil00} we have used the undepleted pump and small signal approximation ($|a_p|^2 \gg |a_s|^2 \gg |a_i|^2$) which explains the asymmetric treatment of the signal and idler modes in (\ref{eqn:cmt1}). We have also assumed no nonlinear losses such as two-photon absorption (TPA) and negligible self- and cross-phase modulation (SPM and XPM).  We set the actual input frequencies of the pump and signal ($\omega_p$ and $\omega_s$ respectively)  equal to their corresponding resonant frequencies.  The generated idler frequency is set by energy conservation ($\omega_i = 2\omega_p-\omega_s$) and is not necessarily equal to $\omega_{i,o}$ (i.e. the idler is not assumed to be on resonance allowing for the inclusion of dispersion in the model).

Solving (\ref{eqn:cmt1}) and (\ref{eqn:cmt2}) in the steady state gives the conversion efficiency 
\begin{align}
\eta \equiv {|s_{i,-}|^2\over{|s_{s,+}|^2}} = 
P_p^2 \omega_i^2 |\beta_{\mathrm{fwm},i}|^2 \left [{2r_{p,\mathrm{ext}}\over{(r_{p,o} +r_{p,\mathrm{ext}})^2}}\right ]^2 \times \label{eqn:efficiency}  \\  {2r_{s,\mathrm{ext}}\over{(r_{s,o} +r_{s,\mathrm{ext}})^2}} {2r_{i,\mathrm{ext}}\over{(2\pi \Delta \nu)^2 + (r_{i,o} +r_{i,\mathrm{ext}})^2}} \nonumber
\end{align}
where $\Delta \nu$ is the detuning of the generated idler from resonance (i.e. $2\pi\Delta \nu = 2\omega_{p,o} -\omega_{s,o}- \omega_{i,o}$) and $P_p$ is the pump power in the waveguide ($P_p=|s_{p,+}|^2$). Assuming the FWM coefficient and pump and signal decay rates are independent of $\Delta\nu$ we see that the efficiency takes on a Lorentzian form with respect to idler detuning 
\begin{align}
\eta \propto  {2r_{i,\mathrm{ext}}\over{(2\pi \Delta \nu)^2 + (r_{i,o} +r_{i,\mathrm{ext}})^2}} \label{eqn:efficiency2}  
\end{align}
where $(r_{i,o} +r_{i,\mathrm{ext}})/\pi$ is the linewidth (in Hz) of the idler resonance.

\begin{figure}[t]
  \centering
  \includegraphics[width= \columnwidth]{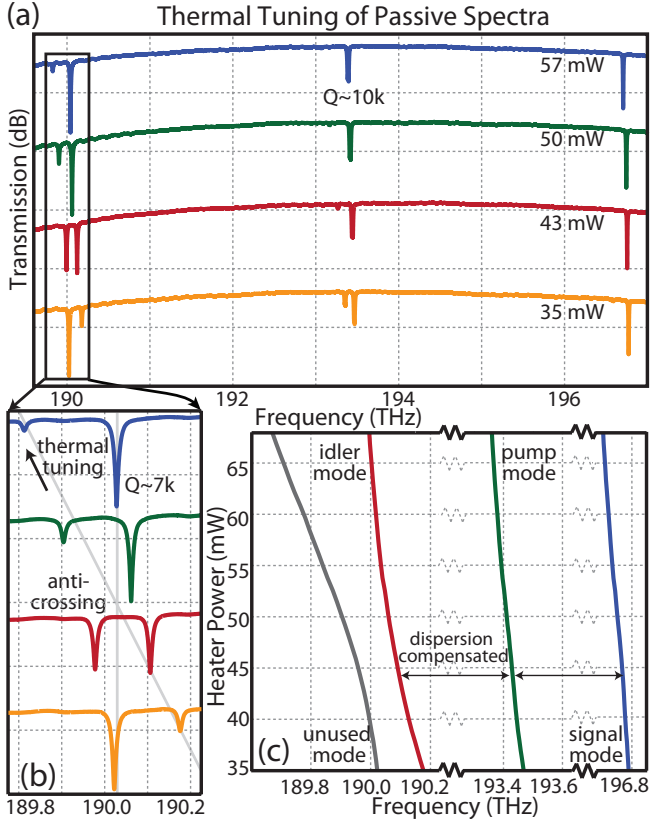}
  \vspace{-25pt}
\caption{(a) Passive spectra with four heater powers with (b) zoomed into the interacting resonances. (c) Tracing of the four relevant resonance frequencies with heater power. \label{fig:fig2}}
\vspace{-15pt}
\end{figure}
We note, specific to our resonator configuration, that the assumption that the FWM coefficient $\beta_{\mathrm{fwm},i}$ is independent of $\Delta \nu$ is not rigorously valid, since the idler mode becomes distributed across both cavities as the dispersion compensating frequency splitting occurs and the mode is more or less hybridized as a result [Fig.~\ref{fig:fig1}(d)].  However, we argue that the FWM coefficient can, at most, be reduced to $1/{\sqrt{2}}$ that of the uncoupled state (assuming maximally split idler supermodes are needed to compensate the dispersion). This is a much slower dependence than the Lorentzian rolloff in equation (\ref{eqn:efficiency2}) and therefore can be ignored for understanding the underlying physics of the system. However, the reduced mode overlap does imply that this form of dispersion compensation can reduce the total efficiency to $1/{2}$ that of a single \textit{dispersionless} resonator of the same size as the primary cavity (while noting that a dispersionless design of the same mode volume and Q may not be possible). This may be a necessary trade-off and small price to pay to gain a small mode volume and large FSR, enabling broadly separated signal and idler and potentially higher overall conversion.
\begin{figure}[b!]
\vspace{-25pt}
  \centering
  \includegraphics[width= \columnwidth]{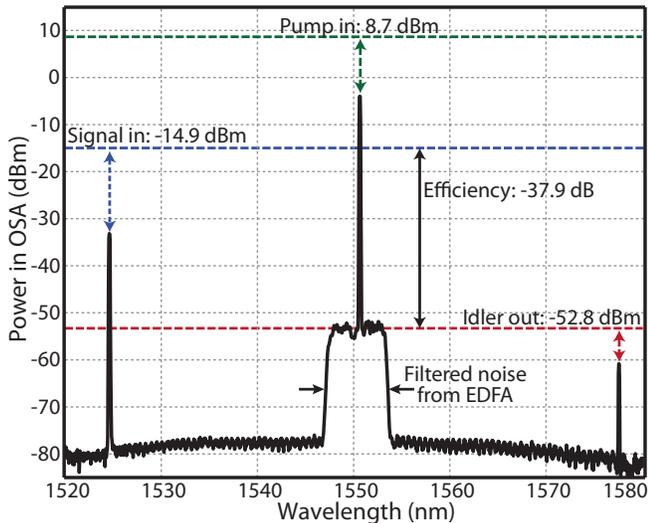}
\vspace{-25pt}
\caption{OSA output spectrum of highest efficiency (-37.9\,dB) seeded four-wave mixing measurement. Dashed lines represent the corresponding powers of each mode in the waveguide. \label{fig:fig3}} \vspace{-10pt}
\end{figure}

\begin{figure}[t!]
  \centering
  \includegraphics[width= \columnwidth]{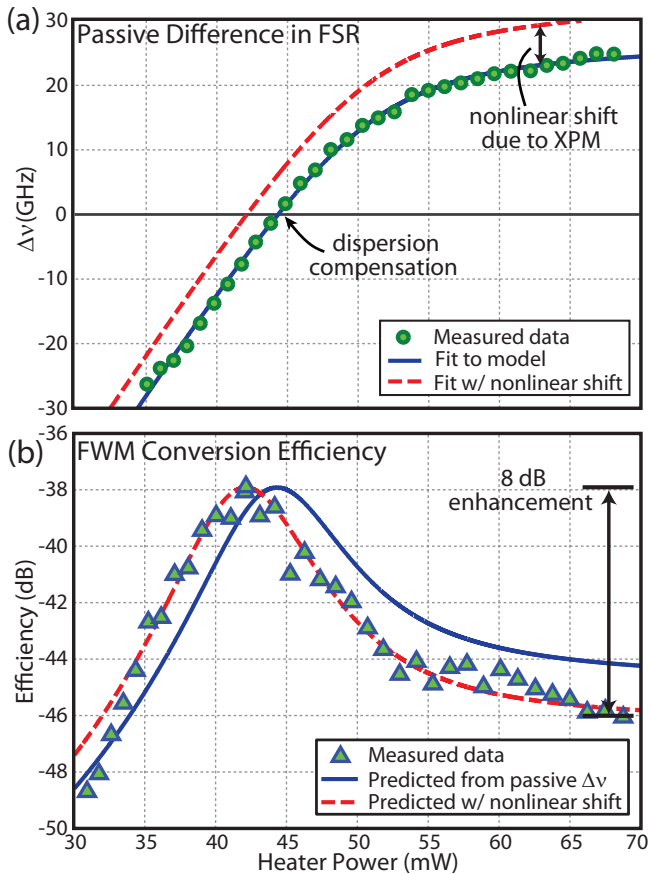}
\vspace{-25pt}
\caption{(a) The passive frequency mismatch measured and fit (solid line) between adjacent FSRs demonstrating dispersion compensation at 44.3\,mW. The dashed line includes estimated nonlinear shift of 6\,GHz due to phase modulation. (b) Measured FWM efficiency versus heater power with normalized (to maximum experimental efficiency) theoretical curve for the shifted (dashed) and non-shifted (solid) fit values in (a) in eq. (\ref{eqn:efficiency2}) \label{fig:fig4}}
\vspace{-20pt}
\end{figure}

The concept was experimentally demonstrated in a silicon microring system fabricated in a silicon-on-insulator platform with a 220\,nm thick silicon device layer, a 2\,$\mu$m buried oxide, and a 1\,$\mu$m overcladding.  The fabricated device consists of a primary and auxiliary ring with 3.5\,$\mu$m and 3.61\,$\mu$m radius respectively [Fig.~\ref{fig:fig1}(c)].  The primary ring has 27\,GHz frequency mismatch due to dispersion. The device includes an additional auxiliary waveguide bus which is very weakly coupled for device characterization. A resistive heater for thermal tuning was fabricated above the auxiliary cavity on the silica overcladding using contact photolithography.  Fig.~\ref{fig:fig2}(a,b) shows the passive (low-power) spectral responses of the through port of the primary ring with thermal tuning.  Maximum mode splitting occurs at a heater power of 43\,mW.  The heater power at which optimum dispersion compensation ($\Delta\nu = 0$) occurs was found to be 44.3\,mW [Fig.~\ref{fig:fig4}(a)]. Seeded FWM was then demonstrated at a range of heater powers spanning from 30 to 70\,mW as shown in Fig.~\ref{fig:fig4}(b). At each heater power, the pump and signal wavelengths were subsequently tuned to optimize conversion efficiency to account for slight changes in the absolute resonance frequency of each mode due to thermal crosstalk and temperature variations. To achieve high pump power, a tunable telecom single frequency laser was amplified by an erbium-doped fiber amplifier (EDFA) followed by two cascaded 5\,nm wide band-pass filters to remove most of the EDFA's amplified spontaneous emission noise. Each measurement was performed with an estimated pump power of 8.7\,dBm and signal power of $-14.9$\,dBm in the waveguide bus. A peak FWM efficiency of $-37.9$\,dB, corresponding to a generated idler power of $-52.8$\,dBm in the waveguide, was found at a heater power of 42.1\,mW. The output to the optical spectrum analyzer (OSA) for this particular result is shown in Fig.~\ref{fig:fig3}.  The device exhibits dispersion compensation leading to 8\,dB of enhancement of the FWM efficiency from the uncoupled case as shown in Fig.~\ref{fig:fig4}(b). Due to the aforementioned thermal crosstalk of the heater, the wavelengths of the pump, signal, and idler varied by $\sim$1\,nm across the tuning range, centered around 1550\,nm, 1524\,nm, and 1578\,nm respectively. This corresponds to FWM across an FSR of 3.334\,THz ($\sim$27\,nm) which is, to the best of our knowledge, the largest FSR in a silicon resonator demonstrated to support FWM.

We note that the maximum FWM efficiency was found to be at a heater power different than one would predict from the passive spectra (42.1\,mW vs 44.3\,mW). The efficiency dependence predicted from the measured idler linewidth and passively measured detuning ($\Delta\nu$) shown in Fig.~\ref{fig:fig4}(a) according to equation (\ref{eqn:efficiency2}), normalized to the $-37.9$\,dB peak efficiency, is shown by the dashed red line in Fig.~\ref{fig:fig4}(b).  We claim that this is caused by an approximately 6\,GHz shift in $\Delta\nu$ due to XPM at the higher pump powers used during FWM. It is well known that XPM is a factor of 2 larger than SPM due to wavelength degeneracy \cite{Agrawal06}. This effectively shifts the signal and idler modes further down in frequency than the pump mode causing additional FSR mismatch to that caused by dispersion. The green dashed line in Fig.~\ref{fig:fig4}(a) shows the estimated $\Delta\nu$ with the additional assumed 6\,GHz shift resulting in a much better fit to the measured data as shown by the solid blue line in Fig.~\ref{fig:fig4}(b).

In summary, we have proposed a mode coupling approach to dispersion compensation and phase matching in resonant four-wave mixing systems.  We demonstrated resonant four-wave mixing across the largest FSR to our knowledge in silicon using a proposed dual-cavity device designed to compensate dispersion through structural degrees of freedom.  With improvement in Q's and coupling losses, we believe this approach can lead to high efficiency, low-power broadly separated FWM and pair generation. We note also that the dispersion control demonstrated in this paper has multiple additional advantages. For instance, in applications such as heralded photon sources where additional signal-idler modes are undesirable, our configuration can be used to avoid generating extra comb lines. Compensating only one resonance will result in all other FSRs not being energy matched and therefore prevent additional signal-idler modes.  On the contrary, this device can benefit the generation of a comb in an already dispersionless cavity by compensating the deleterious effect of differential self- and cross-phase modulation on the pump mode in an otherwise equispaced comb.  This would address an important problem in comb design.  In addition to shifting resonances, the proposed method also enables a degree of freedom for linewidth engineering \cite{Zeng14}.  By employing mode splitting at the pump resonance, one can design a coupling-engineered correlated photon pair source. Since the pump mode would be distributed across both the primary and auxiliary cavities it would display half the external coupling of the signal-idler modes, allowing critically coupling of the pump mode (to achieve maximum pump field enhancement) and over-coupling of the signal-idler modes, intrinsically improving photon pair extraction efficiencies. 

This work was supported by a Packard Fellowship for Science and Engineering and a CU-NIST Measurement Science and Engineering (MSE) Fellowship.  Photonic devices were fabricated at IMEC, Belgium through the ePIXfab Multi-Project Wafer silicon photonics shuttle runs. Micro-heaters were fabricated at the Colorado Nanofabrication Laboratory (CNL) at the University of Colorado Boulder.


\end{document}